\documentclass[iop]{emulateapj} 

\usepackage{color,graphicx}
\usepackage{amsmath,amssymb}
\usepackage{tabularx}

\newcommand{\iso}[2]{\hbox{${}^{#1}{\rm #2}$}}
\newcommand{\Msun}{\ensuremath{{M}_{\sun}}}

\newcommand{\Rsun}{\ensuremath{{\rm R_{\sun}}}}

\shorttitle{Dust in R Cor Bor stars}
\shortauthors{Karakas et al.}

\begin{document}


\title{R Coronae Borealis Stars are Viable Factories of Pre-solar Grains}


\author{Amanda I. Karakas\altaffilmark{1,2}, Ashley J. Ruiter\altaffilmark{1,3}, 
Melanie Hampel\altaffilmark{1,4}}
\email{amanda.karakas@anu.edu.au}



\altaffiltext{1}{Research School of Astronomy and Astrophysics, 
  Australian National University, Canberra, ACT 2611, Australia}
\altaffiltext{2}{Kavli IPMU (WPI), The University of Tokyo, Japan}
\altaffiltext{3}{ARC Centre of Excellence for All-Sky Astrophysics (CAASTRO)}
\altaffiltext{4}{Argelander-Institut f\"{u}r Astronomie, University of Bonn, 
Auf dem H\"{u}gel 71, D-53121 Bonn, Germany}


\begin{abstract}
We present a new theoretical estimate for the birthrate of R Coronae Borealis (RCB) stars
that is in agreement with recent observational data. 
We find the current Galactic birthrate of RCB stars to be $\approx$ 25\% 
of the Galactic rate of Type Ia supernovae, assuming that RCB stars are 
formed through the merger of carbon-oxygen and helium-rich white dwarfs. 
Our new RCB birthrate ($1.8 \times 10^{-3}$ yr$^{-1}$) is a factor of 
10 lower than previous theoretical estimates. This results in roughly 180--540 
RCB stars in the Galaxy, depending on the RCB lifetime.
From the theoretical and observational estimates, we calculate the total dust 
production from RCB stars and compare this rate to dust production from novae and 
born-again asymptotic giant branch (AGB) stars. 
We find that the amount of dust produced by RCB stars is comparable to the amounts
produced by novae or born-again post-AGB stars, indicating that these merger
objects are a viable source of carbonaceous pre-solar grains in the Galaxy.
There are graphite grains with carbon and oxygen isotopic ratios consistent with 
the observed composition of RCB stars, adding weight to the suggestion that
these rare objects are a source of stardust grains.
\end{abstract}






\section{Introduction}

R Coronae Borealis (RCB) stars are a rare class of hydrogen-deficient supergiants with 
atmospheres composed primarily of helium and carbon, where 
the ratio of carbon atoms to oxygen atoms exceeds unity \citep[C/O $>$1,][]{clayton96}.
They also show  abrupt changes to their apparent visual magnitude.
RCB stars are sometimes observed at their maximum light but they have also been
observed during their decline phase, in some cases showing 
sudden declines of up to 8 magnitudes in a few weeks. The star then typically
stays faint for a period of several months \citep{clayton12}
before eventually recovering back to maximum brightness over a
period of several months or longer.  The steep decline in brightness is due to dust 
clouds which are ejected from the star and cover the photosphere when they are 
in the line of sight. Observations indicate that this dust is made of amorphous carbon 
grains which cover a range of sizes up to tenths of a micron 
\citep{garcia11,garcia13b,jeffers12}.

The number of RCB stars known in the Galaxy is about 100 \citep{tisserand13}, 
and more are being discovered or will be discovered in on-going and upcoming surveys of the Galaxy 
(e.g., WISE, GAIA, Euclid, LSST and SkyMapper). It is relatively easy to find RCB stars
because they are variable and are bright, with absolute magnitudes between 
$-5.2 \le M_{\rm V} \le -3.4$, where the faint limit can be extended to 
$-2.6$ \citep{tisserand12}. To identify the object as RCB the typical decline in
the light curve has to be identified but the process is simplified because the
minimum is very conspicuous in the light curve.  Previous estimates suggest that
the number of RCB stars in the Galaxy is higher, at over a 1000 \citep[e.g.,][]{clayton12}.
In \S\ref{sec:rcb} we  discuss the observations, formation scenarios and theoretical 
models of RCB stars in more detail.

Given that the known number of Galactic RCB stars has increased, we want to test whether
these objects could be a viable source for producing carbonaceous dust in the
Galaxy and if it would be possible to find their nucleosynthesis signature 
in pre-solar dust grains.
Pre-solar grains are minerals that survived the formation of the Solar System
  and can be found in primitive meteorites. Different stellar origins
have been proposed for pre-solar grains including AGB stars, supernovae, novae, and 
post-AGB stars.  The composition of
pre-solar grains varies greatly and indicates many different types of grains and stellar
origins \citep[e.g.,][]{zinner98,lodders05,zinner14}. The bulk of pre-solar grains, both
carbon and oxygen-rich types, are associated with the ejecta from 
supernovae or AGB stars. However, a few grains have
been associated with rarer sources such as nova outbursts \citep{amari01b,jose04} 
and post-AGB stars \citep{jadhav13}. \citet{amari01c} briefly discussed RCB stars
as a source for silicon carbide (SiC) grains of type A$+$B, which comprise about
4-5\% of the measured SiC grains to date and are characterized by their carbon
isotope ratios, which show \iso{12}C/\iso{13}C $<10$. \citet{amari01c} discarded
the idea because the majority of carbon isotope measurements for RCB stars 
are \iso{12}C/\iso{13}C~$> 40$ as we discuss in \S\ref{sec:rcb}. Furthermore,
SiC grains are not seen around RCB stars \citep{lambert01,garcia11}.

In order to test these ideas we need to know the numbers of RCB stars relative
to novae and born-again post-AGB stars (see \S\ref{sec:dust}).
However, of importance for dust
production is the dust ejection mass and the lifetime so we also need estimates of
these quantities, which we obtain from theory or observations.
This paper is organised as follows. We first provide an overview
of the observations and theory of RCB stars in \S\ref{sec:rcb}.
In \S\ref{sec:model} we use the results from binary population synthesis models 
to obtain the number of RCB stars expected from theoretical models where we assume 
that all RCB stars are formed by double white dwarf (WD) mergers. In \S\ref{sec:dust}
we take observational and theoretical estimates for the number of RCB stars to provide 
a dust-production rate. We also perform a similar calculation for novae 
and post-AGB stars. We finish in \S\ref{sec:conclude} with a discussion of our
results.

\section{The nature and origin of R Cor Bor stars} \label{sec:rcb}

There are two main competing theories for the formation of RCB stars: 1) the
final helium shell flash scenario or 2) the double degenerate WD merger
scenario. The first scenario involves a WD central star of a 
planetary nebulae to undergo a final thermal pulse, which ignites the helium
shell and causes the star to expand to giant dimensions \citep{clayton96}.
While there is evidence for late thermal pulses in the general population
of post-AGB stars \citep[e.g., Sakurai's Object and 
V605 Aquilae,][]{duerbeck96,clayton97,asplund97},
there are serious problems with this scenario explaining the lifetime of RCB stars.
We refer to Table~3 in \citet{clayton12} for an overview of the
double degenerate versus final flash formation scenarios.
 
The second scenario states that RCB stars are the merger products of carbon-oxygen (CO)
WD and helium-rich WD pairs \citep{webbink84}. The progenitor binary system 
experienced at least one phase of common envelope (CE) evolution (see \S\ref{sec:model}).
Due to radiation of gravitational waves the system
undergoes orbital decay which causes the two stars to approach each other and finally
coalesce. During this merger, the less massive helium WD gets completely disrupted. The
matter from the helium WD is believed to form a Keplerian disc which gets assimilated onto
the surface of the accretor, where it starts to burn \citep{jeffery11}.
The remnant of the helium WD might also form an extended envelope around this merging
product \citep{clayton11}. 

RCB stars have pulsation periods on the order of 40 to 
100 days, from which pulsation masses of $\approx 0.8-0.9\Msun$ have been derived \citep{clayton96}.
Models have also shown that the merger of a CO WD and a helium WD is predicted to 
result in a mass of about $0.96\pm 0.13\Msun$ \citep{han98}, which is consistent with 
the pulsation masses \citep{clayton96,clayton11}. 
It has also been hypothesized that such mergers may lead to thermonuclear explosions such 
as Type Ia supernovae (SNe~Ia), but this remains a topic of investigation 
\citep[see e.g.][]{dan14,pakmor13}.
\citet{han98} theoretically predicted the birthrate of RCB stars to be $1.8\times 10^{-2}$ yr$^{-1}$. 
In \S\ref{sec:model} we update the predicted RCB birthrate for the Galaxy.  

\citet{jeffery11} presented surface abundances of RCB with respect to
evolutionary models which assume that RCB stars are formed by a double degenerate
merger of a CO WD and a helium WD.  The models were compared to the observations
\citep{asplund00,rao03,rao08,pandey08}, which reveal that RCB stars are enhanced in several elements 
such as nitrogen, fluorine, sodium, aluminium, phosphorus, silicon, sulfur, and 
in some neutron capture elements produced by the $slow$ neutron capture process.
Observations also show unexpectedly low isotopic ratios of \iso{16}O/\iso{18}O
with ratios close to or less than unity \citep{clayton05,clayton07,garcia10}. The ratios
of \iso{12}C/\iso{13}C are generally high, with a lower limit of 30 to 40
\citep{cottrell82,hema12}. There are exceptions, including V CrA which has a low \iso{12}C/\iso{13}C
ratio of $\approx 3$ to 4 \citep{rao08}, and VZ Sgr with 3--6 \citep{hema12}.

The elemental and isotopic abundances of RCB stars are reasonably well matched
by merger models, such as those by \citet{jeffery11}, \citet{longland11} and \citet{menon13}.
The models by \citet{jeffery11} assume most of the observed surface composition
of RCB stars comes from the nucleosynthesis from the AGB phase of the CO WD.  
\citet{jeffery11} identified the need for additional nucleosynthesis during the 
merger to explain the composition of some elements, such as fluorine, which shows 
abundances higher than predicted.  
\citet{menon13} reproduces the abundances of C, N, O and F using a one-dimensional 
post-merger evolution and nucleosynthesis model based on realistic hydrodynamic 
merger progenitor models. However, the large Si and S abundances observed in the low
metallicity RCB stars are not reproduced, indicating that some high temperature burning at 
$T \approx 10^{9}$K may be occurring.

\section{Rates of R Cor Bor stars from population synthesis} \label{sec:model}

Theoretical (or extrapolated) estimates for the number of RCB stars 
range from 200 up to 5,700 \citep{webbink84,lawson90,han98,clayton12}.
To estimate the number of RCB stars expected to currently be in the Galaxy, we calculate
theoretical merger rates between helium-rich and CO WDs (hereafter He-CO mergers) using the 
binary evolution population synthesis code {\sc StarTrack} \citep{belczynski02,belczynski08}.

Recently, {\sc StarTrack} input physics have been updated to include a new prescription
for accretion of helium-rich material on CO and ONe WDs when stable Roche lobe overflow
is encountered \citep{ruiter14}. The updated prescription does not affect double WDs that 
merge upon contact (like RCB progenitors), but does affect binaries involving He-CO systems 
that do not merge upon contact and thus lead to AM CVn systems, `classical' double-detonation 
Type Ia supernovae, or other transient events. 
We use the simulations from the P-MDS model of \citet{ruiter14} to calculate the birthrate 
of RCB stars arising from He-CO mergers. All stars are evolved from the ZAMS, and the evolutionary 
histories are recorded until $t=13.7$ Gyr is reached \citep[see][Section~2, for a description 
of initial separations, initial mass function, etc.]{ruiter09}.  For these simulations we 
have adopted a ZAMS binary fraction of 70\% and a burst star formation history, where the 
burst occurs at $t=0$. A `burst' simulation is the most powerful since it allows us to easily 
convolve our event rates (delay times; see below) with any star formation rate of choice 
to attain a realistic model star formation history.

We note that the delay time distribution\footnote{The delay time distribution (or DTD) 
is the distribution of events in time -- this case, mergers -- that occur following a 
hypothetical burst of star formation at $t=0$. The DTD shape is a useful tool in quantifying 
progenitor ages for explosive stellar phenomena, e.g. Type Ia supernovae.} shape of He-CO 
mergers relative to CO-CO mergers is qualitatively similar starting from 500 Myr post-starburst,
owing to the fact that similar physical processes play a role in determining the 
post-nuclear burning orbital evolution of double WDs.
CO-He mergers do not occur with delay times $< 500$ Myr due to the low-mass nature of the
secondary star: A helium-WD progenitor takes longer to complete nuclear evolution than a
CO-WD progenitor.

Given the similar DTD shapes beyond 500 Myr, it is reasonable to assume a constant birthrate of RCB
to  CO-CO mergers at the current epoch (${\sim} 11$ Gyr). 
We find the number of He-CO mergers averaged over a Hubble time to be
$2.65 \times 10^{-14}$ \Msun$^{-1} {\rm yr}^{-1}$, where the mass represents mass born in stars. 
This value is 25 \% of our CO-CO merger rate:
$1.06 \times 10^{-13}$ \Msun$^{-1} {\rm yr}^{-1}$.
This rate is remarkably similar to the rate of 
Type Ia supernovae in Sbc-like galaxies such as the Milky Way: $1.1 \times 10^{-13}$ \Msun$^{-1} {\rm yr}^{-1}$, 
or $7 \times 10^{-3}$ SNe~Ia yr$^{-1}$ as cited by \citet{badenes12} when adopting a Milky Way 
stellar mass of $6.4 \pm 0.6 \times 10^{10}$ \Msun \citep{macmillan11}\footnote{Note that our 
simulations are normalized per unit mass born in stars where as the observational value from 
\citet{badenes12} is normalized per unit stellar mass in a given galaxy. We have not attempted 
to correct for this difference since it unnecessarily introduces uncertainties into our rate. 
However, changing the normalization technique to more closely match that of observations would 
lead to a slight increase in our rate, thus we expect even better agreement.}.
We adopt this Galactic SN~Ia rate in order to extrapolate our RCB population to Galactic numbers: 
we take the specific rate of He-CO mergers to the observed specific Galactic SN~Ia rate which yields
an RCB birthrate of $1.8\times 10^{-3}$ RCB
stars yr$^{-1}$.

\begin{figure}
  \begin{center}
    \includegraphics[width=0.98\columnwidth]{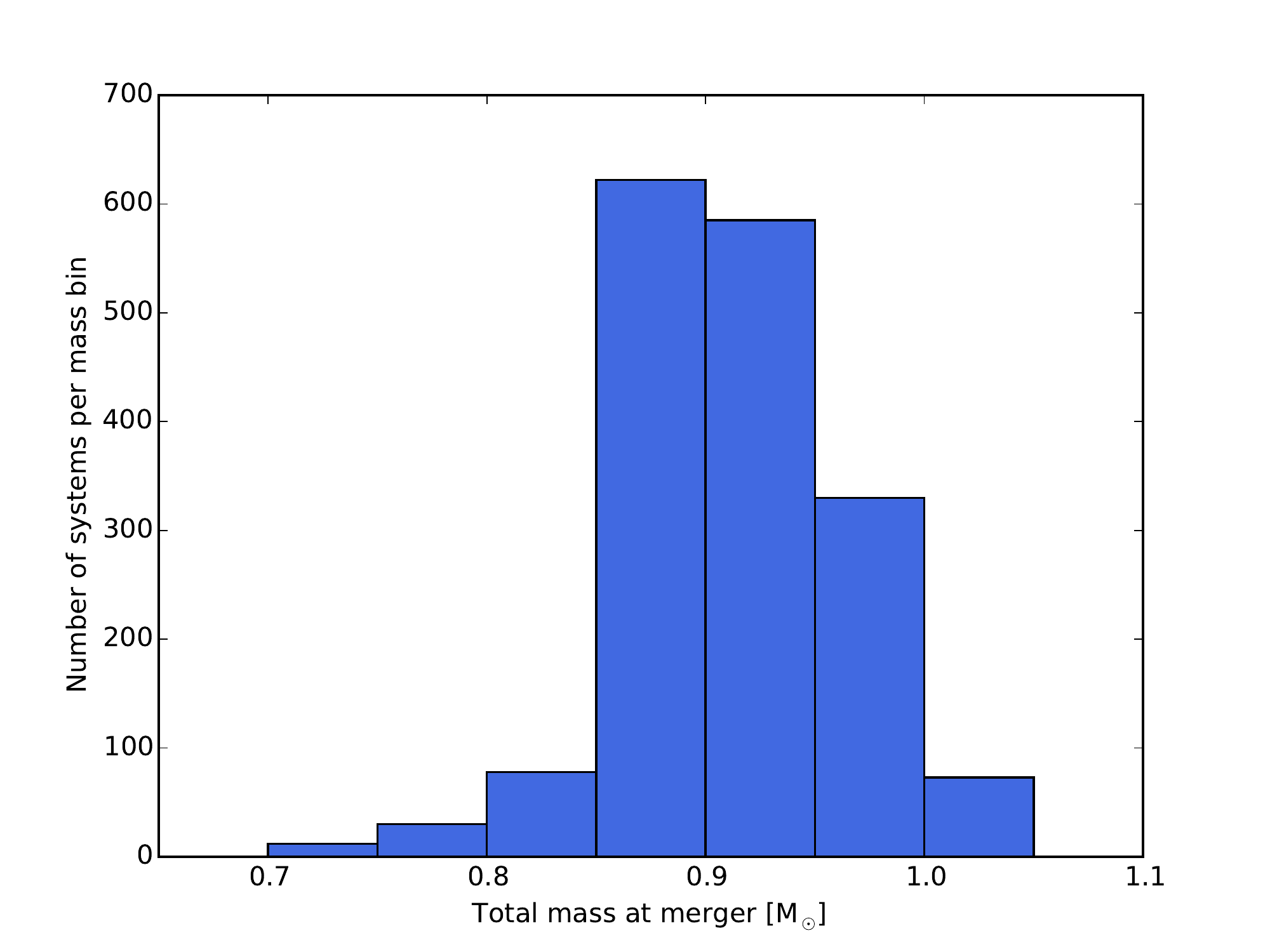}
    \caption{Total mass of RCB systems at time of merger grouped into mass bins of $0.05$ \Msun. 
  We assume all RCB stars are formed through the merger of a CO WD and a helium-rich WD.}
    \label{fig:mtot}
  \end{center}
\end{figure}

In Figure~\ref{fig:mtot} we show the distribution of total masses for He-CO mergers 
over a Hubble time calculated with {\sc StarTrack}. The mass distribution peaks at $\approx 0.9\Msun$, 
which is consistent with the theoretical estimates of \citet{han98} as well as the inferred RCB 
masses derived from pulsation studies \citep{clayton96,clayton11}. We note that in principle
observationally-measured RCB masses are likely to be equal to or less than these theoretically 
predicted RCB `birth' masses, because the merger product will likely lose some mass over time.

We find that RCB stars are formed via three main evolutionary channels, all of which involve 
one CE event. Factors such as ZAMS mass and initial orbital separation play a role in how soon 
after star formation the He-CO merger occurs, as well as what the component masses and compositions 
will be at time of merger. Two of the evolutionary channels (hereafter channel 1 and channel 2) 
involve the merger of a CO WD and a `hybrid' helium-rich WD, where the WD has a CO-rich core and 
a helium-rich mantle. Such HeCO hybrid WDs can be formed in cases where a red giant star is stripped 
of its envelope during binary interactions, and the helium core only undergoes partial burning
\citep{tutukov96}. The third evolutionary channel (channel 3) involves the merger of a CO WD and a 
He WD, the latter which never underwent any helium core burning.

The main difference between channels 1 and 2 is that the ZAMS masses are typically larger by ${\sim}1\Msun$ 
for both components in channel 1. Channel 1 CE events occur when the secondary is a red giant, and the 
WD mergers occur relatively quickly after star formation (within 0.5$-$1 Gyr). For channel 2, the CE events 
occur while the secondary is an early AGB star, and the WD mergers occur 1$-$4 Gyr post-star formation. 
Channel 3 binary components are slightly less massive both on the ZAMS and at time of merger. The 
secondary star is found on the early AGB during the CE phase, and the WD mergers occur $> 4$ 
Gyr after star formation.

\begin{figure}
    \begin{center}
          \includegraphics[width=0.95\columnwidth]{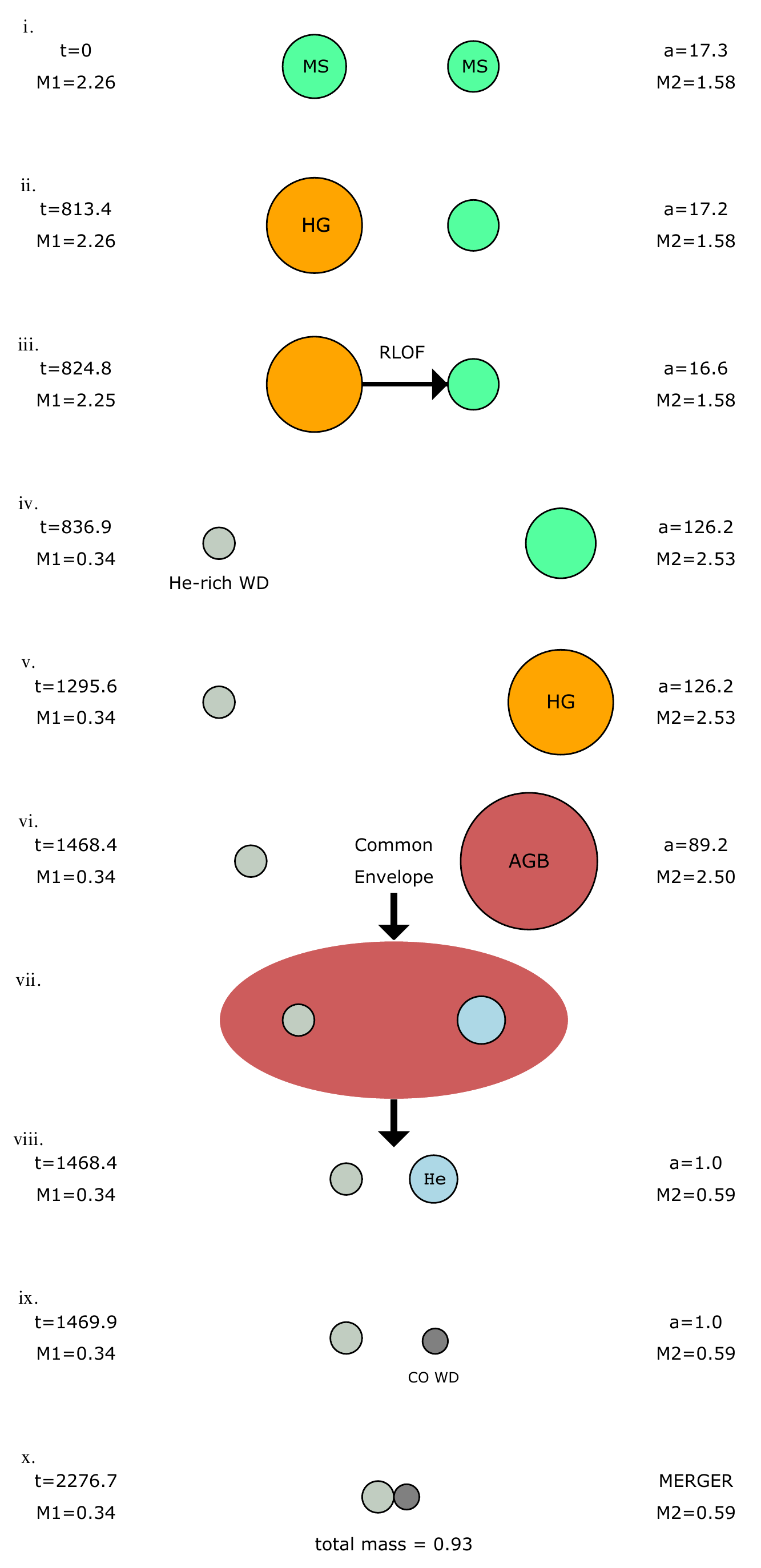}
\caption{Evolution showing one of the three principle evolutionary channels through which RCB 
stars are expected to form, as predicted by our {\sc StarTrack} model. Only the ten most interesting 
phases of evolution are shown (see text). Times are in Myr since birth on ZAMS, separations ($a$) 
are in \Rsun, and masses ($M1, M2$) are in \Msun. The phase of stellar evolution prior to WD 
formation is denoted as follows: MS = Main Sequence, HG = Hertzsprung Gap, AGB = Asymptotic Giant 
Branch, He = (stripped envelope) helium-burning star.}
    \label{fig:evol}
  \end{center}
\end{figure}

We find the most common formation scenario to be of channel 2 type, followed closely by 
channel 3. In Figure~\ref{fig:evol} we show a typical evolution for an RCB star formed via channel 2 
from our population synthesis model. Other than birth on the ZAMS and the time of merger 
(marked i and x in the figure, respectively), we have depicted stages where either a stellar 
component first evolves off the main sequence (ii, v), initiates a mass transfer phase (iii, vi-viii), 
or becomes a white dwarf (iv, ix). 
Over the course of evolution, the binary undergoes a stable mass transfer episode when the primary 
star is on the Hertzsprung Gap, which follows into the red giant phase. After the primary star evolves 
into a He-rich (hybrid) WD, a CE event occurs while the secondary is an early AGB star. The CE leaves 
behind a naked helium-burning star in a relatively close orbit (1 \Rsun) with the He-rich WD. After 
the second star becomes a WD, it takes ${\sim} 800$ Myr for the stars to be brought into contact 
under the influence of gravitational wave radiation. 

\section{Dust production rates} \label{sec:dust}

An important factor when determining the dust production rates from various sources
is not just the number of them at any one time in the Galaxy, but also the amount of
dust that each source produces. In this section we present estimates for the 
number of RCB stars, novae and born-again post-AGB stars and their dust-production rates.
Table~\ref{tab:dust} is a summary of our results, where we highlight the source,
the birthrate, the dust mass formed per event, and the dust production rate. 

\begin{table*}
  \caption{Birthrates, expelled dust mass and 
dust production from RCB stars, novae and born-again post-AGB stars in the Galaxy.}
  \label{tab:dust}
\begin{tabularx}{\linewidth}{l X X X} \hline \hline
      Source & Birthrate [yr$^{-1}$] & Dust expelled per event [$\Msun$] & 
Dust production rate [$\Msun$ yr$^{-1}$] \\ \hline
      R Coronae Borealis stars & $1.8\times 10^{-3}$ & $0.1-1 \times 10^{-6}$  & $1 - 50 \times 10^{-4}$ \\
      Novae explosions & $1-10 \times 10^{1}$ & $10^{-10}$ to $10^{-5}$ &  $10^{-8}$ -- $10^{-3}$ \\
      Born-again post-AGB stars & $0.63-28\times 10^{-1}$ & $4 \times 10^{-6}$ & $7.5 - 34 \times 10^{-6}$ \\ \hline
\end{tabularx}
\end{table*}

\subsection{R Coronae Borealis stars}

In the last decade the observed number of RCB stars in the Galaxy has risen 
from ${\sim} 40$ \citep{zaniewski05} to 76 \citep{tisserand13}.
\citet{lawson90} extrapolated from the number of known RCB stars at the time to 
estimate that there are 200 to 1000 RCB (and cool hydrogen deficient carbon) 
stars in the Galaxy, whereas \citet{clayton12} extrapolated from the number of known RCB 
stars in the LMC to arrive at a considerably higher number of 5,700 RCB stars.

We perform a similar extrapolation here. \citet{tisserand09} find 22 RCB stars in the Magellanic 
Clouds, with 18 in the Large Magellanic Cloud (LMC) and 4 in the Small Magellanic Cloud (SMC).
The Milky Way Galaxy is roughly 10 times more massive than the LMC and 100
times more than the SMC. If we crudely assume that the number of objects scale with galaxy mass
(P. R. Wood, private communication), we estimate that there are between 180 and 
400 RCB stars in our Galaxy. Only part of the LMC has been searched by MACHO
and OGLE and that the number of RCB stars in that galaxy may be significantly higher than 18.
Either way, our new numbers are significantly less than estimated by \citet{clayton12}.
Our new theoretical birthrate of $1.8\times 10^{-3}$ RCB stars yr$^{-1}$, a factor of
10 lower than predicted by \citet{han98},  results in 
$\approx 180-540$ Galactic sources when taking RCB lifetimes between $1-3 \times 10^{5}$ years.
This theoretical estimate is in good agreement with our extrapolated number of RCB stars.

Observational estimates of the amount of carbon-rich 
dust ejected from RCB stars ranges from $10^{-7}\Msun$ yr$^{-1}$ \citep{clayton92} to
$10^{-6}\Msun$ yr$^{-1}$ \citep{feast86}. These numbers are consistent with the 
latest observational estimate of $9 \times 10^{-7}\Msun$ yr$^{-1}$ from \citet{clayton11} 
and \citet{jeffers12}. In Table~\ref{tab:dust} we calculate our dust-production
rates using $10^{-6}\Msun$ yr$^{-1}$; choosing the lower limit of  $10^{-7}\Msun$ yr$^{-1}$ 
would lower our dust production rates by an order of magnitude. 
Furthermore, we have assumed that all RCB stars produce the same amount of 
carbon-rich dust, which may not always be the case.

Taking the observed number of RCB stars in the Galaxy to be $\sim 100$
we obtain a dust production rate of $M_{\rm dust}^{\rm RCB} \approx 10^{-4} \Msun \, {\rm yr}^{-1}$ , 
assuming a dust mass-loss rate of $10^{-6}\Msun$ yr$^{-1}$. Using our newly calculated RCB birthrate, 
adopting the upper limit for the RCB lifetime of $3 \times 10^{5}$ years, 
and again assuming a dust mass-loss rate of $10^{-6}\Msun$ yr$^{-1}$, 
our best theoretical estimate yields a dust production rate of
$M_{\rm dust}^{\rm RCB} \approx 5.4 \times 10^{-4} \Msun$ ${\rm yr}^{-1}$.
In order to find a hard upper limit, we take the theoretical birthrate 
from \citet{han98} of $1.8\times 10^{-2}$ yr$^{-1}$, an RCB lifetime of $3 \times 10^{5}$ years which
results in 5400 RCB stars. This yields a dust production rate of $5 \times 10^{-3}\Msun$ yr$^{-1}$. 
Our range of dust-production rates for RCB stars is summarized in Table~\ref{tab:dust}.

One substantial uncertainty in this estimate is the lifetime
of RCB stars. That there are only about 100 found out of the $\approx 200-5000$ theoretically 
predicted may simply reflect a shorter lifetime for the RCB phase than estimated by
e.g., \citet{clayton12}. 

\subsection{Novae}

We now compare these rates to the dust production expected from novae.
Similar to the situation for RCB stars, the frequencies and mass-loss rates for novae
cover a wide range. Galactic nova rate estimates have yielded numbers from
10 yr$^{-1}$ \citep{ciardullo90} to $\approx 100$ yr$^{-1}$ \citep{liller87,shafter97}.
Most estimates are somewhere in the middle of this range at values of around 
$\approx 30-40$ yr$^{-1}$ \citep{hatano97,nelson04,darnley06}. Nova rates are 
better constrained in M~31 than in the Milky Way, with an estimate of 65 yr$^{-1}$
\citep{darnley06}.

The error estimates are highly uncertain. We take the upper limit of $100$ yr$^{-1}$
to ensure that we are not underestimating dust production from novae.
Typical dust formation masses span a wide range from $10^{-10}\Msun$ to an upper 
limit of $\approx 10^{-5}\Msun$, where most of the dust produced by novae is carbon 
rich  \citep{gehrz98}. We will take the lower and upper limits (see Table~\ref{tab:dust})
which yield nova dust production rates of
$M_{\rm dust}^{\rm novae} \approx 10^{-8} \,\,\, {\rm to} \,\,\, 10^{-3} \Msun$ ${\rm yr}^{-1}$.

Note that nova explosions do not always result in detectable dust production, with 
3 out of the 25 sources in \citet{gehrz98} showing no dust, and another 4 with
no information on the types of dust formed.

\subsection{Born-again post-AGB stars}

There is no expected dust production associated with the post-AGB phase because the
stars are too hot to form dust \citep{vanwinckel03}. The dust around 
post-AGB stars is recycled AGB dust. The late and very late thermal pulse post-AGB stars are 
different: These objects are born-again giant stars such as Sakurai's Object that may produce
their own, C-rich dust \citep{duerbeck96}. Born-again stars have been considered a 
source of pre-solar grains \citep{jadhav08,jadhav13}. Few born-again stars have been 
studied so we do not have a good estimate of the observed mass-loss rates. Sakurai's Object 
is the best studied and \citet{vanhoof07} calculate a lower limit to the mass of 
the total ejecta to be $6 \times 10^{-4}\Msun$. The most recent estimate gives
a dust mass of $4 \times 10^{-6}\Msun$ (A. Zijlstra, private communication). We assume 
that each born-again AGB star ejects the same amount of dust as Sakurai's Object over 
the born-again lifetime. The lifetime of a born-again AGB stars is approximately that
of a thermal pulse  ($\tau_{TP} \approx 10^{2}$ years), giving an average 
dust mass-loss rate of $4 \times 10^{-8} \Msun$ yr$^{-1}$. 

While we do not know the number of born-again AGB stars
in the Galaxy, we can estimate how many there are using the number of post-AGB stars.
\citet{kamath14} and \citet{kamath15} recently estimated the number of post-AGB stars
in the LMC and SMC to be 75 and 34, respectively. These numbers include both
single and binary post-AGB stars, and they derive a combined post-AGB lifetime of 
$\approx 5\times 10^{3}$ years, consistent with lifetimes from evolutionary tracks
\citep{vw94}. If we scale the number of post-AGB stars in the Magellanic Clouds to 
the Galaxy like we did for RCB stars we obtain between 750--3400 post-AGB stars. 

To calculate the birthrate of post-AGB stars in the Galaxy we require a lifetime,
which has been theoretically estimated to be $\approx 10^{3} - 10^{4}$ \citep{vw94}.
We obtain 0.18 to 3.4 yr$^{-1}$.   The lower values are consistent with estimates 
of planetary nebula birthrates of about 0.5 yr$^{-1}$ by \citet{zijlstra91}, whereas 
\citet{moe06} find a higher value of $1.7 \pm 0.3$ yr$^{-1}$ for the entire Milky Way. 
\citet{moe06} also give the post-AGB birthrate of $2.2 \pm 0.5$ yr$^{-1}$, which is estimated 
to be 90\% of the white dwarf birthrate. The post-AGB birthrate from \citet{moe06}
is closer to our upper estimate of 3 yr$^{-1}$.

\citet{iben84} and \citet{renzini82} estimate that 10--25\% of all stars that leave 
the AGB experience a late thermal pulse. Using our new extrapolation for the number
of post-AGB stars in the Galaxy, and assuming that 25\% of all 
post-AGB stars experience a late thermal pulse, we obtain between 188 to 850
born-again AGB stars.  Using the expelled mass provided
in Table~\ref{tab:dust} we estimate that born-AGB stars produce
$M_{\rm dust}^{\rm born-again} \approx 7.5-34 \times 10^{-6} \Msun$ ${\rm yr}^{-1}$.

Assuming a lifetime of $10^{2}$ years for a late thermal pulse
allows us to estimate the birthrate of born-again AGB stars to be between 1.9--8.5 yr$^{-1}$. 
The upper limit seems unreasonably high. Either we have overestimated the number of post-AGB
stars or we have underestimated the lifetime of a late helium-shell flash.
The low-mass ($\lesssim 2\Msun$) AGB models of \citet{karakas14b} spend a 
few times $10^{2}$ years with a convective He-shell.  In Table~\ref{tab:dust} we 
assume the timescale of a He-shell flash is 300 years, which gives birthrates of 
$0.63-2.83$ born-again post-AGB stars per year. These birthrates are much more reasonable 
when compared to the post-AGB birthrate from \citet{moe06}. 

Unlike RCB stars, not all of the dust from born-again AGB stars may be carbon rich but this is 
yet to be determined. Reducing the number of post-AGB stars that experience late thermal pulses 
to 10\% lowers the number of stars to 75--340, and the amount of dust produced 
to $3-14\times 10^{-6}\Msun$ yr$^{-1}$. Note that increasing the duration of thermal pulse
to 300 years will decrease the dust mass-loss rate.

\section{Discussion and concluding remarks} \label{sec:conclude}

We present a new theoretical birthrate for RCB stars of $1.8\times 10^{-3}$ yr$^{-1}$,
a factor of 10 lower than the previous estimate by \citet{han98}.
We note that in the simulations of \citet{han98}, it was assumed that all 
double degenerates that reach contact will eventually merge (Z. Han, private 
communication, 2015). Therefore, one would naturally expect a higher RCB 
birthrate than we obtain here, since in {\sc StarTrack} we do not assume 
all double degenerates merge. Whether a double degenerate binary merges 
depends on the adopted mass transfer stability criteria for a given model 
\citep[see][]{toonen14}, and will strongly impact the resulting merger rates.
\citet{shen15} recently suggested that every interacting double white 
dwarf binary may merge, which would increase the predicted number of RCB stars.
Using our new birthrate we predict that there should be roughly 180--540 RCB stars 
in the Galaxy. We also estimate the number of RCB stars independently from the number 
known in the Magellanic Clouds, and arrive at an 
estimate consistent with our new theoretical predictions. 

We calculate the dust production rate from RCB stars, novae and born-again post-AGB
stars with our results presented in Table~\ref{tab:dust}.
We show that the dust mass-loss rate from RCB stars is comparable to or higher than
novae and born-again post-AGB stars, even when accounting for the large uncertainty 
in the numbers of these objects in the Galaxy today. Our simple arguments along 
with updated observational and theoretical estimates for the numbers of RCB stars 
in the Galaxy show that RCB stars are a viable source of carbonaceous pre-solar grains.  

Are there any pre-solar grains that show the chemical signature of RCB stars? Recall that 
RCB stars have high C/O ratios, \iso{12}C/\iso{13}C ratios $\gtrsim 40$ and \iso{16}O/\iso{18}O 
ratios between $\approx 0.2-20$ \citep{clayton07,garcia10}, and some show the signature of 
heavy element production by the $s$ process \citep{jeffery11,menon13}.
\citet{garcia10} also attempt to derive \iso{14}N/\iso{15}N ratios but only provide
lower limits due to the non-detection of the \iso{12}C\iso{15}N line.

The pre-solar grain database from \citet{hynes09} has a comprehensive list of all the measured
isotopic and elemental compositions of pre-solar grains from the literature. Here we focus on the
graphite grains as these likely formed in a highly \iso{12}C-rich environment.  Furthermore,
silicon carbide features have not been detected around any RCB stars to date \citep{garcia11,garcia13b}.
\citet{amari93} observe that graphite grains with the largest
\iso{18}O excesses also have the highest \iso{12}C/\iso{13}C ratios. These could be associated
with an RCB origin. There are roughly 1800 graphite grains with carbon isotopic ratios that
match RCB stars but only about 10 grains with measured \iso{16}O/\iso{18}O ratios $< 25$. All 10
grains have large \iso{12}C/\iso{13}C ratios ($\ge 160$) and high \iso{26}Al/\iso{27}Al
ratios $\ge 0.028$, consistent with hydrogen burning. \citet{menon13} predict that
Al can be produced, depending on the interplay between the temperature and the burning
timescale but they do not provide predictions for \iso{26}Al in particular.  They also
note that both Si and Ca is observed enhanced in RCB stars but their models do not reproduce
these abundances.  Neutron captures  also occur in the models by \citet{menon13}, which could 
explain the excesses of \iso{29,30}Si observed in the graphite grains. Unfortunately they 
do not provide isotopic predictions for comparison to the grain data. Future work is needed 
to further explore the viability of whether the grains with excesses in \iso{18}O could 
indeed have originated from RCB stars.

\acknowledgments

The authors thank the anonymous referee for comments that helped improve the clarity of the paper.
The authors thank Albert Zijlstra, Devika Kamath, Peter Wood, and Hans van Winckel for informative 
discussions about dust production and post-AGB stars. We also thank Patrick Tisserand for helpful 
information on RCB stars and Brian Schmidt for general discussion. We are grateful to Geoff Clayton 
who provided helpful comments on this manuscript. We also thank Zhanwen Han for discussions 
on double degenerate mergers. 
AIK is grateful for the support of the NCI National  Facility at the ANU, and was supported through an
Australian Research Council Future Fellowship (FT110100475).
 AJR is thankful for funding provided by the Australian Research Council Centre of
Excellence for All-sky Astrophysics (CAASTRO) through project number
CE110001020.

\bibliographystyle{apj}
\bibliography{apj-jour,library}


\end{document}